\documentclass[preprint,prl,showpacs,amsmath]{revtex4}

\usepackage{graphicx}
\usepackage{dcolumn}
\usepackage{bm}

\newcommand{\ltwid}{\mathrel{\raise.3ex\hbox{$<$\kern-.75em\lower1ex\hbox{$\sim$}}}}
\newcommand{\gtwid}{\mathrel{\raise.3ex\hbox{$>$\kern-.75em\lower1ex\hbox{$\sim$}}}}

\begin{document}

\renewcommand{\thefootnote}{\fnsymbol{footnote}}
\title{The State of the Universe}

\author{James B.~Hartle\footnote{EM:
hartle@cosmic.physics.ucsb.edu}}
\affiliation{Department of Physics, 
University of California\\ 
Santa Barbara, California 93106-9530 USA}

\date{\today}

\begin{abstract}

What is the quantum state of the universe?
That is the central question of quantum
cosmology.  This essay describes the place of that
quantum state in
a final theory governing the regularities
exhibited universally by all physical systems in the
universe.  It is possible that this final theory
consists of two parts: (1) a dynamical
theory such as superstring theory, and (2) a
state of the universe such as Hawking's
no-boundary wave function. Both are
necessary because prediction in quantum
mechanics requires both a Hamiltonian and a
state. Complete ignorance of the state leads
to predictions inconsistent with
observation. The simplicity observed in the
early universe gives hope that there is a
simple, discoverable quantum state of the
universe.  It may be that, like the
dynamical theory, the predictions of the
quantum state for late time, low energy
observations can be summarized by an
effective cosmological theory. That should
not obscure the need for a fundamental basis
for such an effective theory which provides
a unified explanation of its features.  It
could be that there is one principle that
determines {\it both} the dynamical theory
and the quantum state.  That would be a
truly unified final theory.

\end{abstract}

\maketitle


\section{Introduction}

The universe has a quantum state.  What is
it? That is the central question of quantum
cosmology --- the subject to which Stephen
Hawking has contributed so many seminal
ideas.

To ask this question is to assume that the
universe is a quantum mechanical system. We
perhaps have little direct evidence of
peculiarly quantum mechanical phenomena on
large and even familiar scales, but there is
no evidence that the phenomena that we do
see cannot be described in quantum
mechanical terms and explained by quantum
mechanical laws. Further, every major
candidate for a fundamental dynamical law
from the standard model to M-theory conforms
to the quantum mechanical framework for
prediction. If this framework
applies to the whole thing, there must be a
quantum state of the universe.

It would be even more interesting if quantum
mechanics broke down on cosmological scales.
But there is not a shred of evidence for
that, and my guess is that, even if it does,
we will only find out by pursuing the
assumption that quantum mechanics is the framework for a
final theory of cosmology.

My talk will not review any of the current
ideas for a quantum state of the universe
even the no-boundary wave function
\cite{NBWF1,11}. The articles of Don Page and
Alex Vilenkin in this volume do that. Rather, I want to
concentrate on explaining why a theory of
the quantum state of the universe must be
part of any final theory. 
I am also not going to discuss the
generalizations of usual quantum theory that
are required for quantum cosmology
\cite{Str01}.  For the essential points of
this talk you can just imagine that the
universe is a vast number of particles in a
very large expanding box. 

\section{Final Theories}

The final theory (to borrow Steve Weinberg's
term) predicts the regularities that are
exhibited by all physical systems ---
without exception, without qualification,
and without approximation.  Much of this
conference has been concerned with the
search for the final theory. A possible view
at present is that it consists of two parts:
\begin{itemize}

\item A universal dynamical law such as
string theory or its successors;

\item A law for the quantum state of the
universe such as Hawking's no-boundary wave
function of the universe.
\end{itemize}

In the model universe in a box these two ingredients
are the Hamiltonian specifying the form
of the Schr\"odinger equation
\begin{equation}
itr\ \frac{d|\Psi(t)\rangle}{dt} = H
\, |\Psi(t)\rangle
\label{one}
\end{equation}
and the initial quantum state
\begin{equation}
|\Psi(0)\rangle
\label{two}
\end{equation}
with which it starts off.

Both of these pieces are necessary for
prediction.  The Schr\"odinger equation by
itself makes no predictions. Probabilities
$p_\alpha$ in quantum mechanics for a
set of alternatives represented by
projection operators $\{P_\alpha\}$ are
given by
\begin{equation}
p_\alpha = \Vert P_\alpha |\Psi (t\rangle
\Vert^2
\label{three}
\end{equation}
and to compute these the quantum state is
needed at least at one time. No state, no
predictions.

To put the matter in a different way, if the
state is arbitrary, the predictions are
arbitrary. Pick any probabilities $p_\alpha$
you like for the alternatives $P_\alpha$.
There is some state that will reproduce
them. For example, you can take
\begin{equation}
|\Psi (t) \rangle = \sum\nolimits_\alpha
p^{\frac{1}{2}}_\alpha |\Psi_\alpha \rangle
\label{four}
\end{equation}
where the $|\Psi_\alpha\rangle$ are any set
of eigenstates of the $P_\alpha$'s
\begin{equation}
P_\alpha |\Psi_\beta\rangle =
\delta_{\alpha\beta} |\Psi_\beta\rangle.
\label{five}
\end{equation}
The $|\Psi (t)\rangle$ constructed according
to (\ref{twofour}) will reproduce the
pre-assigned probabilities $p_\alpha$ in
(\ref{twothree}). 

Neither is ignorance bliss. If you assume
you know nothing about the state of the
universe in a box then you should make
predictions with a density matrix
proportional to unity
\begin{equation}
\rho=\frac{I}{Tr(I)}
\label{six}
\end{equation}
reflecting that ignorance. But this density
matrix corresponds to equilibrium at
infinite temperature and its predictions are
nothing like the universe we live in. In
particular, there would be no evolution
since $[H, \rho]=0$. There would be no
second law of thermodynamics since the
entropy $-Tr[\rho \log \rho]$ is already at
its maximum possible value.  There would be
no classical behavior since, although the
expected value of a field averaged over a
spacetime volume $R$ might be finite, its
fluctuations, $\langle \phi (R)^2\rangle$,
would be infinite. 

The search for a unified fundamental
dynamical law has been seriously under way
at least since the time of Newton with string
theory or its generalizations being the most
actively investigated direction today. By
contrast, the search for a theory of the
quantum state of the universe has only been
actively under way since the time of
Hawking, let us say on this occasion.  Why
this difference?

Dynamical laws govern regularities in time
and its an empirical fact that the basic
dynamical laws are local in space on scales
above the Planck length.  The laws that
govern regularities in time across the whole
universe are therefore discoverable and
testable in laboratories on earth. By
contrast many of the regularities predicted
with near certainty by the quantum state
occur mostly in space on large cosmological
scales.  Only recently has there been enough
data to confront theory with observation.
That difference in the nature of the
predicted regularities, or their difference
in scales, should not obscure the fact that
the state is just as much a part of the
final theory as is its Hamiltonian.

Given these differences, what grounds do we
have to hope that we can discover the
quantum state of the universe? There are
two: The first is the simplicity of the
early universe revealed by observation ---
more homogeneous, more isotropic, more
nearly in thermal equilibrium than the
universe is today. It's therefore possible
that the universe has a simple, discoverable
initial quantum state and that all of the
complex universe of galaxies, stars,
planets, and life today arose from quantum
accidents that have happened since and the
action of gravitational attraction. The
second reason is the idea that the quantum
state and the dynamical theory may be
naturally connected as in Hawking's
no-boundary theory.

\section{Effective Theories}

We are used to the idea of effective
dynamical theories that accurately describe
limited ranges of phenomena. The
Navier-Stokes equations, non-relativistic
quantum mechanics, general relativity,
quantum electrodynamics, and the standard
model of particle physics are all familiar
examples. To construct an effective theory
we typically assume a coarse-grained
description (restricting attention to
energies below the Planck scale for
instance) and assume some simple property 
that the state might predict there (classical
spacetime, for example).

Cosmology too has its effective theories and
its standard model. This is summarized
neatly by Martin Rees in his cosmologists'
credo \cite{Ree01}.  I reproduce it here with
unauthorized additions.
\begin{itemize}
\item spacetime is classical,
\item our universe is expanding,
\item from a hot big bang,
\item in which light elements were
synthesized,
\item there was a period of inflation,
\item which led to a flat universe today,
\item structure was seeded by gaussian
irregularities,
\item which are relics of quantum
fluctuations,
\item the dominant matter is cold and dark,
\item but a cosmological constant (or
quintessence) is dynamically dominant.
\end{itemize}
Possibly all current observations in
cosmology, at least the large scale ones,
can be compressed into this list of ten
assumptions and a few cosmological
parameters. That is not unlike the situation
in particle physics where most observations
can be compressed into the Lagrangian of the
standard model and its eighteen or so
parameters.

However, the success of such effective
theories which operate in limited ranges of
phenomena should not obscure the need to
find fundamental ones which apply to all
phenomena without qualification and without
approximation. It would be inconsistent, I
believe, to pursue a fundamental dynamical
theory in the face of a successful standard
model, and not pursue a fundamental theory
of the state of the universe because of the
success of its standard model. That not
least because the fundamental theory could
provide a unified explanation of its
assumptions.

It must be said, however, that when the
natural domains of fundamental theories are
as far from controllable experiments as
string theory and the quantum initial
condition the possibility of definitive
tests seems to recede.  It could be that the
predictions of string theory are limited to
general relativity, gauge theories,
supersymmetry, and the parameters of the
standard particle model.  In a similar way
the predictions of the state of the universe
could be limited to classical spacetime, the
initial conditions for inflation, and the
quantum fluctuations that satisfy large
scale structure. Perhaps that is prediction
enough.

\section{Directions}

The instructions of the organizers were to
discuss ``future directions in theoretical
physics and cosmology''. Continuing the
search for a final theory incorporating
dynamics and the initial quantum state is
certainly one direction. But I would like to
mention three questions that might lead to
different approaches to the main one.

\subsection{What's Environmental?}

Which features of the observed universe
follow entirely from the dynamical theory
$(H)$ and which follow entirely from the
initial condition $(|\Psi (0)\rangle)$, and
which are the result of quantum accidents
that occurred  over the course of the
universe's history with probabilities
specified by the combination of $H$ and
$|\Psi (0)\rangle$. Those that depend
significantly on $|\Psi (0)\rangle$ are
called ``environmental''. Some version of
this question was number one on the list of
top ten questions for the next millenium 
prepared by string theorists at the Strings 2000
conference.

Take the coupling constants in effective
dynamical theories for instance. The
viscosities and equation-of-state in the
Navier-Stokes equation are certainly
environmental. They vary with system, place,
and time. But at a given energy do the
coupling constants of the standard model of
the elementary partcle interactions vary
with place and time or with the possible
history of the universe? If so then the
initial quantum state is central to
determining their probabilities.

\subsection{Why Quantum Mechanics?}

The founders of quantum mechanics thought
that the inherent indeterminancy of quantum 
theory ``reflected
the unavoidable interference in measurement
dictated by the magnitude of the quantum of
the action'' \cite{Boh58}. But why then do we
live in a quantum mechanical universe which
by definition is never measured from the
outside?

The most striking general feature of quantum
mechanics is its exact linearity --- the
principle of superposition. But why should
there be a principle of superposition in
quantum cosmology which has only a single
quantum state?

\subsection{Why a Division into Dynamics and
Initial Condition?}

The schema for a final theory which I have been
describing  posits a separate theory of
dynamics and quantum state. Could they be
connected? They already are in Hawking's
no-boundary wave function
\begin{equation}
\Psi \left[h_{ij}, \chi\right] = \int
\delta g\, \delta\phi\ e^{-I[g, \phi]}
\label{seven}
\end{equation}
where the action for metric
$g_{\alpha\beta}(x)$ coupled to matter
$\phi(x)$ determines both the state and
quantum dynamics.  Is there a principle that
determines both? Is there a connection
between superstring theory and its
successors and a unique quantum state?

A unified quantum theory of state and
dynamics would be truly a final theory. That
is surely a direction for theoretical
physics and cosmology.


\section{Acknowledgments}

This paper was supported in part by NSF Grant
\#PHY00-70895.

\end{document}